# The role of time estimation in decreased impatience in Intertemporal Choice

Camila S. Agostino[1,2], Peter M. E. Claessens[2], Fuat Balci[3] and Yossi Zana[2]

[1]Department of Biological Psychology, Faculty of Natural Science, Otto-von-Guericke Universität Magdeburg, Germany, [2]Center for Mathematics, Computing and Cognition, Federal University of ABC, Santo André, Brazil, [3]Departmentof Psychology, Koç University, Istanbul, Turkey

## Author Note

C.S.A had a research scholarship from Capes and Analysis, Imaging, and Modelling of Neuronal and Inflammatory Processes (ABINEP) International Graduate School.

The authors declare that the research was conducted in the absence of any commercial or financial relationships that could be construed as a potential conflict of interest. The Research Ethics Committee at Koç University approved all experimental protocols. All participants signed a consent statement. All authors consented this publication.

All data, code and material can be available upon request.

Correspondence concerning this article should be addressed to Camila S. Agostino, Department of Biological Psychology, Faculty of Natural Science, Otto-von-Guericke Universität Magdeburg, Universitätsplatz, 2, Building 24, room 102, 39104, Magdeburg, Germany. E-mail: camila.agostino@ovgu.de

The role of specific cognitive processes in deviations from constant discounting in intertemporal choice is not well understood. We evaluated decreased impatience in intertemporal choice tasks independent of discounting rate and non-linearity in long-scale time representation; nonlinear time representation was expected to explain inconsistencies in discounting rate. Participants performed temporal magnitude estimation and intertemporal choice tasks. Psychophysical functions for time intervals were estimated by fitting linear and power functions, while discounting functions were estimated by fitting exponential and hyperbolic functions. The temporal magnitude estimates of 65% of the participants were better fit with power functions (mostly compression). 63% of the participants had intertemporal choice patterns corresponding best to hyperbolic functions. Even when the perceptual bias in the temporal magnitude estimations was compensated in the discounting rate computation, the data of 8 out of 14 participants continued exhibiting temporal inconsistency. The results suggest that temporal inconsistency in discounting rate can be explained to different degrees by the bias in temporal representations. Non-linearity in temporal representation and discounting rate should be evaluated on an individual basis.

Decision-making is a central theme to many disciplines such as psychology, economics, neuroscience, and artificial intelligence (Frydman & Camerer, 2016) and inter-temporal choice is one of the key subjects that is common to these fields (Hayden, 2016; Ericson and Laibson, 2018). The task usually involves decisions with consequences that play out over time (Berns et al. 2007). It typically requires choosing between a small reward to be received sooner and a larger one to be received later or the inverse with payments. This task has been investigated in humans and nonhuman animals, such as monkeys (Hwang et al. 2009; Cai et al. 2011; Cromwell et al. 2018), pigeons (Mazur, 1987) and rats (for a review, see Fobbs and Mizumori, 2017). In all cases, there was a tendency to prefer immediate or sooner rewards over delayed ones, with equal or larger values. Interestingly, this behavior could be extensively observed with different kinds of outcomes, meaning that, if an individual prefers sooner, but smaller monetary rewards, he will also prefer sooner and less advantageous primary (food and drink) or other kinds of rewards (for a review, see Odum et al, 2020).

Commonly, if individuals are presented with a pair of choices with rewards to be received immediately or in a near future (e.g. $10 today or $15 in two weeks), the earlier choice tends to be preferred. However, if the same proposal is shifted to a more distant future (e.g. $10 in two weeks or $15 in four weeks), the late reward would most likely be chosen. When a delay is introduced after the initial intertemporal choice, participants tend to reverse their choice. This reversal is also known as dynamical inconsistency (Strotz, 1955; Takahashi, 2005) and it has been extensively demonstrated (e.g., Andreoni & Sprenger, 2012; Sadoff & Samek & Sprenger, 2014; Sprenger, 2015; Augenblick, Nierderle & Sprenger, 2015; Shou Chen & Guangbing Li, 2019). However, how subjective time representation influences inconsistent intertemporal choices still needs further investigation.

## Theoretical Foundation

**Discounting Utility Models**

In standard normative economics, individuals are considered to be rational and aware of their preferences. Early in 1937, Samuelson proposed a model by which individuals behave such that they maximize the sum of future utilities during any specified period of time. The utility of a reward as a function of delay is commonly called the discount utility function. In this discounted-utility model, individuals discount their future preferences over time by a constant rate. When the reward value is decreased by a constant rate, i.e., reward value is consistently discounted in relation to time, the discount function can be described by an exponential equation:

(1) $$\varphi(t) = e^{-\delta t}$$

where $\varphi$ represents the discounted utility function, $t$ is the time interval and $\delta$ is a constant that reflects the discount rate (Samuelson, 1937).

Posteriorly, Strotz (1955) distinguished between the following two conditions: time consistent behavior, in which the choice for any future date is dependent only on the time-distance from the present (Samuelson's model), and time inconsistent behavior, in which preferences might change in the future and the choice depends on the calendar date at which the act will occur. Several studies presented evidence that humans and other species do not discount rewards in an exponential manner (see Frederick et al. 2012 for a review). For instance, one of the first studies that demonstrated this effect was presented by Thaler (1981), who told participants that they won $15 prize in the lottery and could withdraw it from the bank right away. He then asked them how much would be needed for them to accept postponing the withdrawal of the reward. The median values for three month, one year and

three year delays were $30, $60 and $100, respectively. These values reflect, respectively, annual compound discount rates of 277%, 139% and 63%, which is considerably higher for shorter intervals. Similar observations are made with non-human animals. For instance, Mazur (1987) tested pigeons in a task in which they chose between two food rewards with different delays. In his study, he estimated the indifference points for the two delays and found that, for example, two seconds of access to food at a delay of six seconds was equivalent to six seconds of food-access delivered after 17 seconds.

One of the earliest models developed to explain the deviation from constant discounting is the quasi-hyperbolic discount function (Phelps & Pollak, 1968; Laibson, 1997):

(2) $$\varphi(t) = \{y\delta^t, \ t > 0 \ 1, \ t = 0$$

$\delta$ is a constant-rate time-decrement factor with values between 0 and 1, $y$ is a positive constant that represents the "impatience" factor, adding a negative bias to later rewards. Differently from the exponential model, in this model the delayed reward is discounted at a higher rate for earlier than for longer delays, a phenomenon termed future preference reversal or temporal inconsistency (Strotz, 1955; Laibson, 1997). However, from the second period and on, the discount rate is constant.

Based on fitting the data from his work with pigeons, Mazur (1987) proposed a somewhat different hyperbolic model, sometimes called proportional discounting model:

(3) $$\varphi(t) = \frac{1}{(1+\delta t)}$$

where $\delta$ represents a constant associated with the discount rate (larger values represent higher discount rate), thus in this model the resultant discount rate continuously decreases with the time delay.

An alternative, more general, hyperbolic model was proposed by Loewenstein and Prelec (1992).

(4) $$\varphi(t) = (1 + ht)^{\frac{-r}{h}}, h \geq 0, r > 0$$

It can be observed that when $r = h$, the proportional discounting model is obtained. One of the properties of the general hyperbolic formulation is that decreasing impatience, or magnitude of deviation from constant discounting, is represented by the $h$ parameter (Bleichrodt et al. 2016). This is a major advantage because the discount rate and deviation from constant discounting are not confounded. Prelec (2004) defines decreasing impatience as Impatience minus Time-preference and expresses it as $\frac{h}{1+ht}$. This is the definition we adopt in this paper.

**Decreasing Impatience**

The origin of inconsistency in delay discounting was attributed to at least two different cognitive mechanisms. Impulsivity and deficient self-control processes were pointed out as affecting the value of the discounting in non-human animals (Ainslie, 1974) and humans (Loewenstein, 1996), and, more recently, neural evidence associated impatience behavior with a valuation and cognitive control brain networks (Van den Bos & McClure, 2013; O'Connell et al. 2018). Read (2001) illustrated this behavior with a typical impatient dieter individual who promised that, after having a full breakfast, he/she would have a light salad for lunch; at lunchtime, considering the available options at the restaurant, the individual orders fish-and-chips. In a recent study, multi-attribute drift diffusion modelling rejected the hypothesis that individuals integrate monetary and temporal information before the comparison of options, supporting the alternative hypothesis that people compare the two attributes separately (Amasino et al. 2019). Putting together, both evaluations shape the patience factor of an individual's choice. Other authors attributed inconsistency to the subjective temporal representations, which would differ from the actual magnitude of calendar time. Ebert and Prelec (2007) argued that people might be insufficiently sensitive to the duration of time intervals, in part due to lack of attentional focus. However, it was Takahashi (2005, 2006, 2009) who formalized the hypothesis that deviation from normativity in discount functions is due to non-linearity of subjective time.

Empirical evidence in support of this hypothesis was presented by Zauberman and colleagues (Zauberman et al., 2009; Kim & Zauberman, 2009). As a general method, they asked students to perform two tasks: time-interval magnitude estimation and intertemporal monetary trade-offs. A cross-modal matching paradigm was used to estimate how temporal representation changes as a function of objective durations (Epstein & Florentine, 2005). In this method, a time interval is presented and participants mark on a horizontal line such that the marked length best matches the subjective magnitude of the time interval. After the completion of this task, in the intertemporal task, participants were required to indicate how much they would be willing to be paid after a given time interval as compared to an immediate US$ 75 reward. In both tasks, time intervals were in the 3 to 36 months range with steps of 3 months.

The main difference between the Zauberman and colleagues (2009) and Kim and Zauberman (2009) studies is that, in the former, the length of the horizontal line was bounded to 180 mm, while in the latter the line was unbounded. The discount rate was modeled after Mazur's (1987) hyperbolic function. The main assumption was that if a bias in time perception is responsible for the inconsistency in the discounting rate, substituting calendar time by subjective time scale would reduce the inconsistency. The first study (Zauberman et al. 2009) found that the hyperbolic function fit the calendar-based data, but not the data based on the subjective time scale, a result in support of the distorted time perception hypothesis.

However, participants deviated at a surprisingly high level from linearity with respect to calendar time. For example, the 36-months interval was perceived on average as 1.9 (as opposed to 12) times longer than the 3-months interval. The second study (Kim and Zauberman, 2009) did not present the time perception data in the paper, but it was reported that the psychophysical function was less flat as a function of the changes in calendar time and estimated as a power function with an exponent value of .72.

Comparing calendar-based and subjective-time-based discounting rate functions, it was found that, across participants, the discount rate was more constant (i.e. less dependable on the specific time delay) in the latter condition. Additionally, based on individual-level analysis, there was a positive correlation ($r = 0.27$) between the degree of time contraction and hyperbolicity, i.e., individuals that perceived time horizons to be longer overall or perceived time more nonlinearly deviated more from exponential discounting. Han and Takahashi (2012) also showed that the hyperbolic nature of the discounting function is related to the perception of time. They asked participants to perform an intertemporal choice task and estimate time intervals. Based on the q-exponential model and aggregated data across participants, they found that hyperbolicity disappeared ($q > 1$) when the subjective time scale was considered instead of calendar time. It is noteworthy that there were very large gaps between the time interval points (ex. 6 months, 5 years and 25 years time-intervals), implying highly speculative interpolation in the fitting procedures. More recently, Bradford and colleagues (Bradford, Dolan & Galizzi, 2019) replicated the findings by Zauberman and colleagues (2019) by measuring time intervals estimation and time discounting. One of the main contributions of the recent study is in providing further support for the time-perception hypothesis using actual monetary rewards. As demonstrated in previous studies, actual monetary incentives after experimental procedures can affect individuals' motivation and behavior even without conscious awareness. (for review, see Capa & Cluster, 2014).

However, the time estimation and intertemporal choice tasks were not counterbalanced, leaving open the possibility that the estimated discounting function, that requires time estimation, was influenced by the earlier time estimation task. Furthermore, the goodness-of-fit of the estimated subjected time functions were not reported, and lastly, the magnitude of the one-day interval was chosen arbitrarily as a reference to normalize the other intervals estimation.

In spite of the relative prominence of the distorted time perception hypothesis, its empirical support is surprisingly sparse, consisting mainly of evidence from the three cited studies. Other studies investigated temporal inconsistency in intertemporal choice in the time scale of days, but relied on time perception functions estimated from tasks in the seconds and minutes scales. In one study (Brocas et al. 2018), impatience was found to be associated with the subjective experience of slower passage of time. Moreover, data were presented only partially and methodological issues limit the validity of the conclusions. Citing a few aspects, the order of the timing and intertemporal choice tasks was not counterbalanced, the former always preceding the latter. Additionally, the individual-level analysis was attempted, but the experimental design was a between-subjects with no repeated measurements of the participants, which led to reduced reliability of the estimates. Finally, in any single study, only one hyperbolic model was assumed to be right, without demonstrating its better goodness-of-fit as compared to alternative hyperbolic models. The models that were chosen actually made it harder to isolate indices of time inconsistency from the discounting rate.

The aim of the current study is to test the hypothesis that time perception explains, at least in part, the inconsistency in delay discounting, by applying procedures that compensate for the methodological deficiencies of previous studies. We used a direct approach, measuring both temporal magnitude estimation and choice preferences of the same participants at the same time resolution and scale. A repeated measurements design was used with group- and individual-level analyses. Three models were considered: (1) exponential as a constant rate model; (2) proportional hyperbolic as a classic decreasing impatience model; (3) general hyperbolic as a model that allows quantifying discounting rate and decreasing impatience separately.

The main contributions of this study to the fields of behavioral economics and cognitive psychology are both methodological and theoretical. It establishes an experimental design and procedures that can be used to rigorously evaluate intertemporal choice behavior in conjunction with subjective time. Importantly, the statistical analysis approach adopted in this paper covers the spectrum from the purely individual level to the purely group level. Concerning the actual findings, we reproduce the findings that power and hyperbolic functions best fit subjective time and delay discounting, respectively but the study reveals strong group inhomogeneity in both domains, which calls for the necessity of developing theoretical or practical tools to discriminate between individuals as well as to identify the source of inter-subject differences.

## Materials and Methods

All participants were healthy undergraduate students from Koç University (Istanbul, Turkey), who volunteered to take part in the experiments and provided written consent for their participation. Participants received course credits for taking part in the experiment. The Research Ethics Committee at Koç University approved all experimental protocols.

Twenty-six volunteers (ages 19-26, mean 20.76, 23 women) performed two tasks, a temporal magnitude estimation task, and an intertemporal choice task, in counterbalanced order. Two participants were excluded based on their intertemporal task performance[1]. The experimental

---

[1] One participant was excluded due to extreme invariance in her choices, with less than 0.15 difference in discounting rate between the minimum and maximum and a coefficient of variation of 0.05. The responses of another participant were considered outliers: the z-score of the $\beta$ beta parameter of the exponential function fit

protocols for temporal magnitude estimation followed a cross-modality line-length matching paradigm (Zauberman et al., 2009) while the experimental protocol for the intertemporal choice task was adapted from Rodriguez and colleagues (2014). Stimulus presentation and data collection were programmed using MATLAB version R2007 (The MathWorks, Natick, MA) with Psychophysics Toolbox 3 extensions (Brainard, 1997; Pelli, 1997).

## Task I: Temporal Magnitude Estimation

In this task, participants were seated in an isolated laboratory room, at 70 cm from a computer monitor. The following instructions were presented in Turkish on the screen: "In this study, you will be asked to indicate your subjective feeling of durations between today and many days in the future. The days vary between 3 and 36 months. Please, read the instructions carefully and indicate your answer."[2] On the upper part of the screen, a text message instructed the participants to "Imagine the time interval below. Move the bar to indicate how long you consider the duration between today and the given interval."[3] The time interval in months was presented below these instructions in the format "*nn* months" ("*nn Ay*") according to a random permutation of five repetitions from the set {3, 6, 9, 12, 15, 18, 21, 24, 27, 30, 33, 36}. Below the numeric time interval, a 180 mm line (685 pixels) was presented with labels "very short" ("çok kısa") and "very long" ("çok uzun") placed at the left and right extremes, respectively. The initial position of the mouse cursor was always at the center of the line. Participants could move the cursor to the right or left to determine the desired segment length and click the left mouse button to confirm their choice. The maximum

---

was 2.15; the z-score of the h parameter of the hyperbolic function fit was 5.6. See Analysis Section and Table 1 for details and comparison.

[2] "Lütfen talimatları dikkatlice okuyun ve cevabınızı belirtin. Zaman aralıkları 3 ve 36 ay arasında değişecektir. Bu deneyde sizden bugün ve uzak gelecekteki günler arasında geçen süreye ilişkin öznel hislerinizi belirtmeniz istenecektir."

[3] " Aşağıdaki zaman aralığını düşünün. Bugün ile verilen aralık arasındaki sürenin gözünüzde ne kadar olduğunu göstermek için çubuğu hareket ettirin."

response window was 10 seconds, after which a new trial was initiated. No-response trials were treated as missing values in the statistical analysis. Each of the 12 time intervals was presented five times (total of 60 trials per session). Four training trials with a random selection of intervals were presented at the beginning of the task to familiarize participants with the procedure; these data were not included in the analyses.

## Task II: Intertemporal Choice

In this task, an adaptive staircase procedure was adopted. Participants were asked to imagine that they would receive different amounts of money in two different periods of time and that they could choose only one of the alternatives. The presented choices were 100 Turkish Lira (approximately 25 US$) to be received "now" or 150 TL to be received "later" in a time interval in months selected from the set {3, 6, 9, 12, 15, 18, 21, 24, 27, 30, 33, 36}. These intervals were presented in a random sequence in each block of trials. The amount of money associated with each interval of time was adjusted adaptively according to the participant choice: when the "now" alternative was chosen, the next time the same time interval was presented, the value of the "later" reward was increased by 10%. Conversely, when the "later" alternative was selected, the value of the "later" reward was decreased by 10% in the next presentation trial. The choices were made through the computer keyboard, with no time limit. Inter-trial interval was one second and participants were allowed to take a break for about 10 seconds every 30 trials. The experiment was terminated when the participant made at least three choice inversions (Now-Later or Later-Now sequence) after the 10[th] trial for each of the twelve time intervals. The total number of trials over the 12 time intervals was 460 on average.

## Analysis

The psychophysical function of time interval was estimated by fitting a linear or power function to the data while exponential and hyperbolic functions were fit to the intertemporal choice data. Parameter estimation and goodness-of-fit were evaluated using three different approaches. The more traditional methods are the analysis on aggregated data, in which fitting is performed on the average data across all participants, and the two-stage approach, in which functions are fit to the data of each individual and the average of the estimated parameters are taken. In a third multi-level approach, nonlinear mixed-effects (NLME) modeling is used to estimate the function's parameters. In this method, group mean and between-participant variations are considered simultaneously. The final output is influenced by the estimated parameters on the individual and aggregated data, but with particular attention to the quality of the data. For instance, the parameter estimation of individuals whose behavior is more consistent (low variability, fewer outliers, more data points) would have greater weight on the final value of the parameter. The multi-level approach was shown to present advantages over aggregated and two-stage approaches, specifically for estimating discounting functions (Young, 2017). In all cases, fit quality was estimated using the maximum likelihood and intercept, slope and coefficients were allowed to covary across participants.

In the current study, we present results from different approaches for easier comparison with previous studies. Linear functions were of the form $c + \alpha t$, where $t$ is the time interval, and power functions were defined as $c + \alpha t^{\beta}$, thus it is not assumed that the functions intercept the axis at their origin. The Bayesian information criterion (BIC) was used to decide which model better fits the data, where a lower BIC indicates a better model; if the BIC difference was lower than two, the linear model was preferred (Kass & Raftery, 1995).

The discounted value $DV$ as a function of time interval was computed as follows. An equivalence point ($EP$), representing the reward value of indifference between the immediate and delayed options, was defined as the middle point between two trials where an inversion of choice was made ("Later" following "Now" or vice-versa). In this study, all analyses considered the average $EP$ calculated for each time interval as the mean value of the first three inversion points following the 10[th] trial. The discounted value was defined as the proportion of the equivalence point in relation to the "Now" reward, i.e. $DV = \frac{100}{EP}$.

Discounted value results were fit with exponential or hyperbolic functions. The exponential function followed Equation 1. Two functions were used as hyperbolic models: the proportional discounting model (Eq. 3) and the general hyperbolic model (Eq. 4). All analyses were conducted in Matlab and The R Project for Statistical Computing, with the nlme package, Ver. 3.1-137 (Pinheiro et al., 2019).

**Results**

**Temporal Magnitude Estimation**

Figure 1 presents the average line length estimated as a function of time intervals and the best fit linear and power models of 24 participants. At the group level, both linear and power models explained at least 99% of the variance. The BIC value of the linear and power models was 111.6 and 89.7, respectively. The difference (22.9 units) indicates that the power model is preferable to the linear, with a β value of 0.67 (SE±0.04).

The contribution of the between subject variability to the result can be estimated by comparing fitting with a fixed-effects model with and without random effects. A NLME analysis fit by maximum likelihood, with subjects as random effect, favored the power function with a BIC value 51 units lower than that of the linear function (β=0.67, SE±0.04).

The power function was also favored, although with delta BIC of only 15 units, by a nonlinear least squared (NLS) analysis. This analysis considers the same fixed effects as the NLME analysis, but without the random effects responsible for inter-individual variation. This result means that when the inter-subject variability is considered, the power function is even more preferable than when averaged data are used, indicating group inhomogeneity with differential tendency toward power functions time-interval mapping (Young, 2017).

Linear and power models were fit also with the two-stage approach. Averaged $R^2$ was 0.94 (SEM±0.02) and 0.98 (SEM±0.00) for the linear and power models, respectively. The average β value of the power function was 0.70 (SEM±0.06). Out of 24 participants, the data of 16 participants (67%) were fit with the power model with BIC value at least two units lower than the values for the linear model (only in one case the linear model had a BIC value at least two units lower than for the power model). The average β value of these 16 "hyperbolic" participants was 0.57 (SEM±0.08); the low value expresses the frequently observed concave, compressed-shape of the mapping function. Only two participants had β >1.

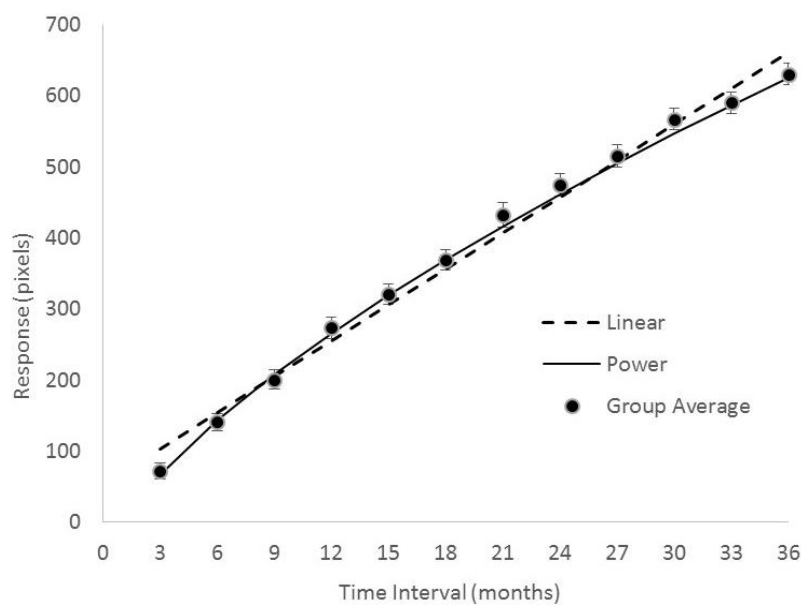

**Figure 1** - *Time intervals magnitude estimation. Dots represent the average length of responses (line segments, in pixels) across all participants. Error bars represent the standard error of the mean. Dashed and dotted lines represent the best (aggregated) fitting linear and power functions, respectively.*

**Intertemporal Choice**

Results from the intertemporal choice task encompass data from the same 24 people that participated in the time interval estimation task (Fig. 2). The exponential function plot (discussed below) helps visualize the inconsistency in the discounting rate, with the data points before and after intervals of 21 months located below or above the curve, respectively. Fitted to the aggregated data, the exponential, proportional hyperbolic and general hyperbolic functions explained 97.1, 97.9, and 97.5% of the variance, respectively (Table 1). BIC values were -24.4, -46.2 and -48.4 respectively, thus the hyperbolic functions are preferable to the exponential. The $\delta$ parameter of the exponential and proportional hyperbolic functions was 0.045 (SE $\pm$ 0.003) and 0.076 (SE $\pm$ 0.002), respectively. The $h$ and $r$ parameters of the general hyperbolic functions were 0.133 (SE $\pm$ 0.032) and 0.094 (SE $\pm$ 0.01), respectively.

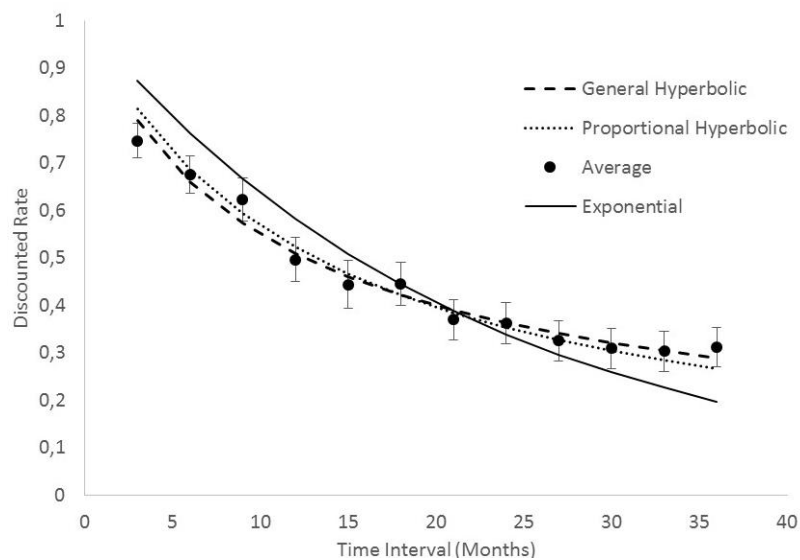

**Figure 2 -** *Discounted values for different time intervals. Each dot represents the average across the 24 participants. Error bars represent the standard error of the mean. Continuous, dotted and dashed lines represent the best fitting exponential, proportional hyperbolic and general hyperbolic functions, respectively.*

NLME model analysis arrived at slightly different results. Both hyperbolic functions were preferable over the exponential, but while the proportional hyperbolic was BIC 58.6 units below that of the exponential, the general hyperbolic was only 3.9 units lower. The $\delta$ parameter of the exponential function was higher (0.060) than in the fixed-effect analysis, as well as the $\delta$ of the proportional hyperbolic function (0.116). The $h$ and $r$ parameters of the general hyperbolic function were also different (0.081 and 0.151, respectively[4]). The influence of the inter-subject variability on the model preference was evaluated, as in the case of the temporal magnitude estimation task, by repeating the data using fixed-effects NLS model. The delta BIC was 26.6 units in favor of the proportional hyperbolic function in comparison to the exponential, lower than the value given by the NLME model analysis. As seen in the time-intervals analysis, the differences in results obtained using fixed-effects and mixed-effects indicate the importance of the analysis of inter-subject variation, in other words, they indicate that group averaging conceals relevant individual differences.

The same three functions were also fit using the two-stage approach. Average $R^2$ was 0.85 (SEM±0.023) and 0.86 (SEM±0.021) and 0.87 (SEM±0.020) for the exponential, proportional and general hyperbolic functions, respectively. Out of 24 participants, the data of 15 participants (63%) were fit with the proportional hyperbolic function with BIC value at least two units below that of the exponential (in 6 cases BIC was at least 2 units lower for the exponential function). When comparing the exponential function and general hyperbolic, 14 participants (58%) had the data better explained by the general hyperbolic function. Temporal inconsistency represents deviation from the exponential model, thus the results mean that for at least 37% of the participants presented no temporal inconsistencies. Interestingly, only 3 out of these 9 participants displayed a linear temporal magnitude estimation.

---

[4] For a sense of the impact of the values of the h and r parameters on the discount rate, the difference between the aggregated and NLME model analysis translates to a gradual increase in difference, with the NLME rate 0.4% (3 months) to 7.6% (36 months) lower.

The hypothesis that a nonlinear time perception explains inconsistencies in discounting rate predicts that the inconsistency observed using calendar time scale is diminished if a subjective time scale is used. One can analytically derive from the generalized hyperbolic function that exponentiating the time parameter by any value between 0 and 1, while keeping t the discounting value constant, would result in a lower $h$ value. However, we preferred to test the hypothesis using the actual data by mapping the calendar time to a time scale of subjective units and repeating the function fitting procedures. The procedure was restricted to individuals whose data were better fit by the general hyperbolic model (n=14). Subjective time was defined as $t^c$, where $t$ is the objective time and $c$ is the exponent of the power function that best fits the time interval estimation data. The modified general hyperbolic function was defined as $(1 + ht^c)^{\frac{-r}{h}}$.

Following remapping of the time scale, the general hyperbolic function explained 97.3% of the variance at the aggregated-level; the average explained variance of the fits to the individual participants was 84.4% (SEM ±2.6 ) (Table 1 and Fig. 3B), thus, the model fit the data reasonably well. Considering the aggregated data analysis, the value of the $h$ parameter decreased from 0.133 before the time-scale modification to 0.031 after the modification, indicating that deviation from consistent temporal discounting was reduced. At the same time, the $r$ parameter increased from 0.094 to 0.133, indicating a relatively minor alteration in the discounting rate. The effect of taking into account the mapping of time intervals on decreasing impatience is illustrated in Figures 3. Considering the objective calendar time scale, the general hyperbolic function fit the data well, while the exponential did not (Fig. 3A). This reflects a decreasing impatience tendency, captured only by the hyperbolic function, with a stronger amplitude at short time intervals (Fig. 3C). When the psychological time interval mapping is considered and time scale adjusted, the data approximate a constant rate discounting function (Fig. 3B), reflecting a more stable decreasing impatience (Fig. 3C).

**Table 1 -** *Functions fit comparisons using objective and subjective time scale. Values refer to two-stage approach level, aggregated-level analysis (across subjects) or NLME analysis of the data. The error term represents standard error.*

| | Exponential | Proportional Hyperbolic | General Hyperbolic | | |
|---|---|---|---|---|---|
| | Objective | Objective | Objective | Objective (n=14 ) | Subjective (n=14) |
| **Two-stage $R^2$** | 0.85±0.023 | 0.86±0.021 | 0.87±0.020 | 0.87±0.021 | 0.844±0.026 |
| **Aggregated $R^2$** | 0.971 | 0.979 | 0.975 | 0.975 | 0.976 |
| *δ* | 0.045±0.003 | - | - | - | - |
| *δ* | - | 0.076±0.002 | - | - | - |
| *h* | - | - | 0.133±0.032 | 0.287±0.065 | 0.147±0.042 |
| *r* | - | - | 0.094±0.010 | 0.137±0.016 | 0.170±0.015 |
| **BIC** | -24.4 | -46.2 | -48.4 | -50.54 | -51.98 |
| **NLME** | | | | | |
| *δ* | 0.060±0.009 | - | - | - | - |
| *δ* | - | 0.116±0.024 | - | - | - |
| h | - | - | 0.151±0.057 | 0.378±0.114 | 0.290±0.118 |
| r | - | - | 0.081±0.005 | 0.134±0.012 | 0.182±0.013 |
| **BIC** | -434 | -492 | -438 | -297 | -292 |

At the individual level, 12 out of 14 participants had a lower *h* parameter after consideration of the subjective time interval mapping. After the time-scale adjustment, eight out of 14 participants remained having their data better explained by general hyperbolic function. These results suggest that subjective mapping of time intervals can explain in part inconsistency in discounting rate. However, further analysis weakens this explanation and points to the role of other factors. We statistically tested the hypothesis that the $h$ parameter of the subjective function (M=0.274±0.532) is lower than the $h$ parameter of the objective functions ($h_{Objective} < h_{Subjective}$) against the alternative hypothesis ($h_{Objective} \geq h_{Subjective}$). A paired-samples Bayesian test found both hypotheses equally acceptable, with BF of 0.965 and 1.036, respectively. Additionally, the time-bias hypothesis expects time intervals mapping to follow a non-linear pattern for participants presenting temporal inconsistency, i.e. hyperbolic discounting function. The results show that 36% of the 14 hyperbolic participants presented a linear time perception mapping that is nearly equivalent to 40% of the non-hyperbolic participants.

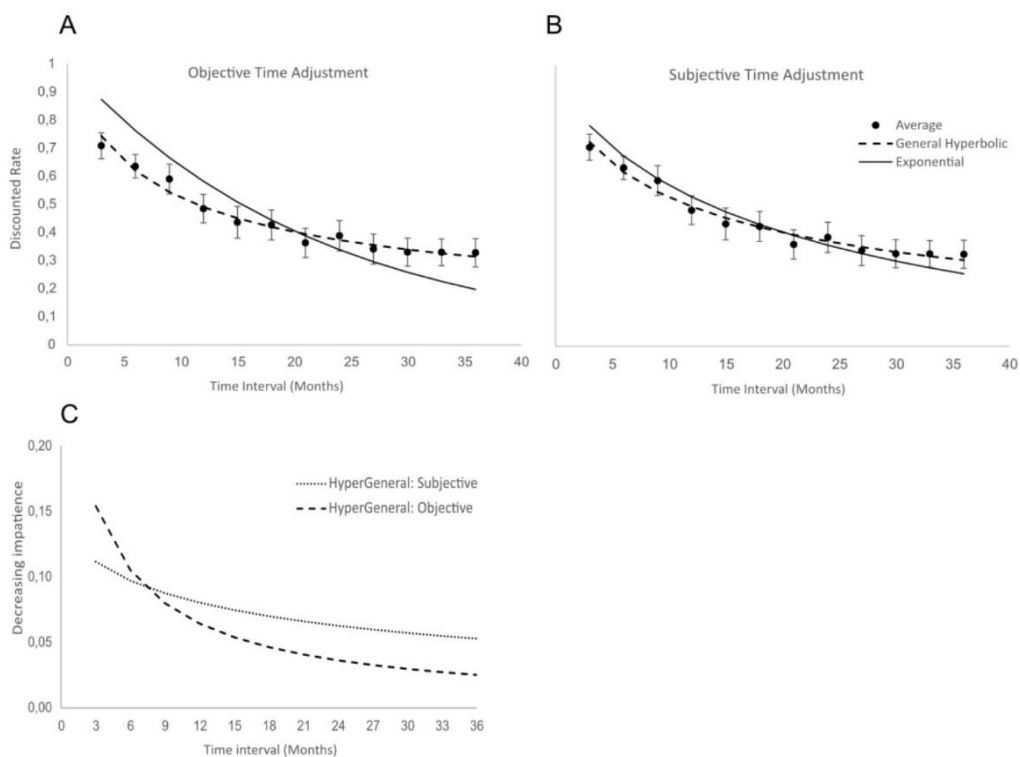

**Figure 3 -** *Panels A and B: Average discounted rate of 14 participants whose data were better fit by the general hyperbolic model. Each dot represents the average across the participants and error bars represent the standard error of the mean. Best fit exponential and general hyperbolic functions are represented by a continuous or dashed line, respectively. The abscissa in Panel A represents the unadjusted objective time interval scale. The abscissa in Panel B represents the time interval scale after adjustment to the time estimation function ($\beta = 0.695$). Panel C: Decreasing impatience estimated from the general hyperbolic function (Prelec, 2004). Thick and thin dashed lines correspond to functions with the objective or adjusted time scale, respectively.*

## Discussion

In most studies of human decision making and intertemporal choice, presentation format and procedure of time intervals is taken for granted as irrelevant for the choice preference, hence they are not considered in the behavioral analysis. However, this procedure invariance premise was not always confirmed. A recent study (Rung et al. 2019) showed that even the progression order of the time interval delays significantly influence the steepness of the discounting function, with the exponential model performing better in the inverse progression presentation condition (i.e., decreasing duration between delays). Still, some investigators focused their attention on the components of the temporal inconsistency in intertemporal choice behavior. The current study presents a rigorous behavioral study of the hypothesis that a cognitive bias in estimation of the magnitude of time intervals can explain the temporal inconsistency. We found that psychometric power functions with exponents between .66 and .67 best explain the temporal magnitude estimation. These functions indicate a compressed-shape, concave psychophysical mapping, implying that people estimate long time intervals in a biased form such that long intervals are estimated as being shorter than estimates made by an objective unbiased observer, i.e., temporal inconsistency. The mapping function based on aggregated data ($\beta$=0.70±0.06) is in the same range of compression previously registered by Kim and Zauberman (2009; $\beta$=0.72) and Agostino and colleagues (2019; $\beta$=0.77).

However, magnitude estimation of time intervals may vary considerably between individuals. In one study, the time estimation results of 40% of the participants were better fit by linear functions (Agostino et al., 2017). The data presented in this study confirm these findings and show considerable inter-subject variability with 33% of the participants presenting data consistent with a linear mapping function. Consequently, the compression level of the psychometric function of the non-linear participants is even higher than the level seen at the group-level analysis. The significance of this phenomenon is that time estimation of some people deviates strongly from calendar time and that any tentative explanation of temporal inconsistency in intertemporal choice tasks needs to account for the individual-level psychophysical time scale mapping. Specifically, it implies that discount functions should be analyzed and discussed at the individual level.

As expected from previous findings (e.g. Ainslie and Haendel, 1983), when evaluated on aggregated-data, hyperbolic functions explained most of the variance of discounted value of future rewards and had stronger empirical evidence in their favor compared to exponential functions. However, individual-level analysis showed that the behavior of 37% of the participants was better explained by exponential functions, thus these participants did not present time inconsistency, i.e. present or future preference. The picture is further grayed out by the finding that two-thirds of individuals in this group displayed temporal magnitude estimation that is better explained by a power function. The between-subjects inhomogeneity in temporal inconsistency was restated by the NLME and NLS analyses results.

These results are consistent with the current body of literature. Individual differences in discounting tasks were found to be associated with several factors, among them substance abuse, gambling, age, psychiatric disorders and cognitive ability (for a review, see Chabris et al. 2008, Basile and Toplak, 2015, Mitschel, 2019). However, notable differences are frequently found also between relatively homogeneous participants. Andreoni and Sprenger (2012) used the β parameter of the quasi-hyperbolic model to estimate present bias. They found that 60% of the participants presented values near 1.0, i.e., no time bias, and 20% displayed a present bias. Ashraf et al. (2006) had similar findings, with 27.5% of the individuals presenting hyperbolic responses and 19.8% were future-biased. Brocas et al. (2018) measured the β parameter of the quasi-hyperbolic model and found a higher proportion of non-biased participants (78%) and a lower proportion of present biased individuals (6%). Sayman and Öncüler (2009) reported that individual differences in temporal inconsistency varied according to the choice options, for example the delay length. Supporting this result, Yoon (2020) found that participants could be distinguished according to their level of patience, and that time consistency peaked at a moderate level of patience.

Individual differences were reported also by Bradford et al. (2019). They found that time intervals are estimated in a non-linear manner consistent with the Weber-Fechner and Stevens' power laws, but none of the personal characteristics could explain the related individual differences. However, the authors asked participants to estimate the duration of the task session and report that 33.7% of them underestimated the time, while the others were accurate. Task-duration underestimation was also associated with a more "expanded" (lower $\beta$) perception of time.The hypothesis that bias in time perception explains individual differences in temporal inconsistency in intertemporal choice tasks is appealing. If the time-interval mapping corresponds to a compressed time estimation (power function, $\beta<1$), hyperbolic individuals could be applying a constant-rate discounting function, i.e., using an exponential temporally-consistent reward devaluation. In the current study, the adjustment of the time scale to account for the non-linearity of the time-interval magnitude estimation reduced time inconsistency such that the data supported more the exponential than the hyperbolic function. However, a representative proportion of the hyperbolic participants had a linear-function time interval mapping, not expected by the temporal inconsistency hypothesis.